# Identification of Iron (III) Peroxo Species in the Active Site of the Superoxide Reductase SOR from *Desulfoarculus baarsii*.

Christelle Mathé[ξ§], Tony A. Mattioli[ξ*], Olivier Horner[¶], Murielle Lombard[§], Jean-Marc Latour[¶], Marc Fontecave[§] and Vincent Nivière[§*]

[ξ] *Laboratoire de Biophysique du Stress Oxydant, SBE/DBJC CEA/SACLAY, 91191 Gif-sur-Yvette Cedex, France.* [¶] *Laboratoire de Physicochimie des Métaux en Biologie FRE 2427, DRDC-CEA/CNRS/Université J. Fourier, CEA-Grenoble, 38054 Grenoble Cedex 9, France.* [§] *Laboratoire de Chimie et Biochimie des Centres Redox Biologiques, DRDC-CEA/CNRS/Université J. Fourier, CEA-Grenoble, 38054 Grenoble Cedex 9, France.*

Superoxide reductase (SOR) is a newly discovered activity by which some anaerobic or microaerophilic organisms eliminate superoxide, $O_2^{\bullet-}$.[1] The SOR catalyzed reaction differs from that of well-known superoxide dismutases SOD in that it does not produce $O_2$, but instead reduces by one electron $O_2^{\bullet-}$ to form $H_2O_2$ exclusively: $O_2^{\bullet-} + 1\ e^- + 2H^+ \rightarrow H_2O_2$.

The active site of SOR consists of a $Fe^{2+}$ center (center II) in an unusual [$His_4$ $Cys_1$] square pyramidal pentacoordination.[2] It reacts specifically at a nearly diffusion-controlled rate with $O_2^{\bullet-}$, generating $H_2O_2$ and the oxidized form of the enzyme, the ferric iron center II. The SORs (originally called desulfoferrodoxin) found in some sulfate reducing bacteria, e.g. *Desulfoarculus baarsii*[1b] and *Desulfovibrio desulfuricans*,[2a,3] contain an additional mononuclear $Fe^{3+}$ center, called center I, coordinated by four cysteines with a distorted rubredoxin-type structure. However, center I is not required for the reaction and, up to now, its function remains unknown.[1b-c]

Recent pulse radiolysis studies of the reaction of center II with $O_2^{\bullet-}$ have allowed the observation, in the micro and millisecond time scale, of intermediates characterized by absorption bands in the 550-650 nm range.[4] These transient species were proposed to be $Fe^{3+}$ peroxo complexes, from which $H_2O_2$ is liberated, on the assumption of an inner sphere mechanism for $O_2^{\bullet-}$ reduction and on the basis that the corresponding absorption bands were slightly different from those of the final ferric iron center II.[4]

On the basis of the crystal structure[2b] and spectroscopic studies[5] of the SOR from *Pyrococcus furiosus*, it has been proposed that upon oxidation the iron active site becomes six-coordinated, as the consequence of a local protein domain movement which places a strictly conserved glutamate (Glu47 in the SOR from *D. baarsii*) in the free coordination site. We have mutated the Glu47 to alanine (E47A) in the SOR from *D. baarsii* and found that this mutation did not affect the kinetics of formation of the above mentioned intermediates detected by pulse radiolysis.[4a-b] However, because this Glu residue becomes a ligand for the oxidized iron, a likely hypothesis could be that it serves to release $H_2O_2$ from the $Fe^{3+}$ peroxo intermediate by substitution in the iron coordination sphere.

Here, we have reacted SOR E47A from *D. baarsii* directly with $H_2O_2$ and have found that the active site of the mutant can indeed transiently stabilize a $Fe^{3+}$ peroxo species, that could be spectroscopically characterized.

When we rapidly manually mixed SOR E47A from *D. baarsii* with 6 equivalents of $H_2O_2$, a UV-visible absorption feature with a maximum at 560 nm, characteristic for the oxidation of the iron center II,[6a] was immediately observed (Fig.1A).[6b] The 4.2 K EPR spectrum,[7] after subtraction of signals from center I, recorded just after addition of 6 equivalents of $H_2O_2$ was complex, with a major feature at g = 4.3 and a minor one at g = 4.15 (Fig.1Bi). The former one is comparable to that of an EPR spectrum of SOR E47A oxidized with hexachloroiridate (IV) (Fig.1Bii). It is characteristic for a high-spin $Fe^{3+}$ in a rhombic ligand field.[1b,3] No other signals in the g = 2 and g = 8-10 regions were observed. At longer incubation time (10 min) with $H_2O_2$, the feature at g = 4.15 completely disappeared (data not shown).

Resonance Raman (RR) spectra at 15 K,[8] taken from the SOR E47A frozen immediately after addition of $H_2O_2$ indicated the presence of two new bands at 850 and 438 $cm^{-1}$ (Fig.2b), which were not present when SOR was oxidized with hexachloroiridate (IV) (Fig.2a). The RR spectra also exhibit a band at 742 $cm^{-1}$ which has been attributed to an internal C-S stretching mode of the CysS-$Fe^{3+}$ active site.[3] When the same Raman measurements were made after mixing with $H_2{}^{18}O_2$, the 850 and 438 $cm^{-1}$ bands were observed to down shift to 802 and 415 $cm^{-1}$, respectively (Fig.2c). RR measurements in $D_2O$ buffer indicated no significant shifts of the 850 and 438 $cm^{-1}$ bands to within 1 $cm^{-1}$ (cf. Supporting Information).

When the reaction was carried out with the wild-type SOR and $H_2O_2$, under the same conditions that we described above for the mutant, an intense RR band at 743 $cm^{-1}$ was observed (Fig. 2d). This band can be used as a marker of the amount of $Fe^{3+}$ formed in these conditions. The bands at 850 and 438 $cm^{-1}$ observed in the case of the mutant with a similar amplitude as that of the 743 $cm^{-1}$ band (Fig. 2b) were now in the case of the wild-type found to be very weak compared to the 743 $cm^{-1}$ band (Fig. 2d). However, they exhibited the same shift upon $^{18}O$ substitution than reported in the case of the mutant (data not shown). The 4.2 K EPR spectra of the SOR wild-type, after subtraction of signal of center I, and recorded immediately after addition of $H_2O_2$, exhibited the rhombic signal at g = 4.3,[1b,3] whereas the feature at g = 4.15 was very weak and completely vanished within a few min (data not shown).

The observed RR frequencies at 850 and 438 $cm^{-1}$ and their $^{18}O$ isotopic shifts (-48 and -23 $cm^{-1}$) are consistent with the $\nu$(O-O) and $\nu$(Fe-$O_2$) stretching modes, respectively, of an $Fe^{3+}$-peroxo species.[9] The lack of deuterium isotopic shifts suggests that this peroxo species is not protonated. We thus conclude that $H_2O_2$ can oxidize SOR and bind to the ferric center II to yield a transient high-spin $Fe^{3+}$-peroxo species, associated with the feature at g = 4.15, as observed from the 4.2 K EPR spectra. The absorption band at 560 nm resulted probably mainly from the Cys-to-$Fe^{3+}$ charge transfer band,[3,5] but also contains a contribution of the peroxo-to-iron $Fe^{3+}$ charge transfer band.[9] The resolution of these two charge transfer bands could be achieved by a RR excitation profile, but this is complicated because of the strong interference of center I when excitations are made below 647 nm.[3]

The observed Raman frequencies are comparable to those described for the end-on high-spin $Fe^{3+}$-OOH species in oxyhemerythrin which showed deuterium isotope shifts.[10] However, for SOR reported here, the unusually low Fe-$O_2$ frequency (438 $cm^{-1}$) strongly suggests a side-on $\eta^2$ $Fe^{3+}$-peroxo species[11] as found in the high-spin Fe complexes such as $[(EDTA)Fe(\eta^2\text{-}O_2)]^{3+}$, for example.[9] In addition, the lack of deuterium shift, suggesting a non-protonated peroxo species, is also consistent with a side-on $\eta^2$ $Fe^{3+}$-peroxo species since it is expected to be more stable in the unprotonated form. Such a coordination in the SOR active site would thus imply either a heptacoordination for the iron or a loss of one of the imidazole ligands, but up to now there is no evidence for such possible coordination changes.[5] Clearly, relevant model Fe-peroxo species with sulfur ligands, not yet available, would support our proposal of a side-on peroxo coordination in SOR.

In conclusion, the data presented here first show that SOR active site can accommodate a $Fe^{3+}$-peroxo species and thus support the hypothesis that reduction of $O_2^{\bullet-}$ proceeds through such intermediates. To our knowledge, this is the first $Fe^{3+}$-(hydro)peroxo species that has been identified in a mononuclear non-heme iron protein, with such an unusual active site. Current RR experiments in the laboratory are directed in order to identify $Fe^{3+}$ -peroxo species formed immediately after reaction with $O_2^{\bullet-}$.

Second, the results suggest that the conserved Glu47 might serve to help $H_2O_2$ release, as illustrated in Scheme 1, since mutation of that residue to alanine results in stabilization of the $Fe^{3+}$ peroxide. It should be noted that the presence of the cysteinate trans to the peroxide may also be crucial in promoting $H_2O_2$ dissociation from the $Fe^{3+}$-peroxo intermediate, by pushing electron density on the iron. As a matter of fact, the Fe-$O_2$ bond observed here, with $\nu = 438$ $cm^{-1}$, is particular weak and the O-O bond with $\nu = 850$ $cm^{-1}$ strong, when compared to the corresponding values reported for model complexes that promote O-O cleavage and formation of high valent Fe-O species.[9]

**Acknowledgement**. TAM thanks P. Mathis and A. W. Rutherford for interest and support in this work. VN and ML thank S. Menage for helpful discussions.

**Supporting Information Available**. Deuterium isotopic effects on the RR bands at 850 and 438 $cm^{-1}$.

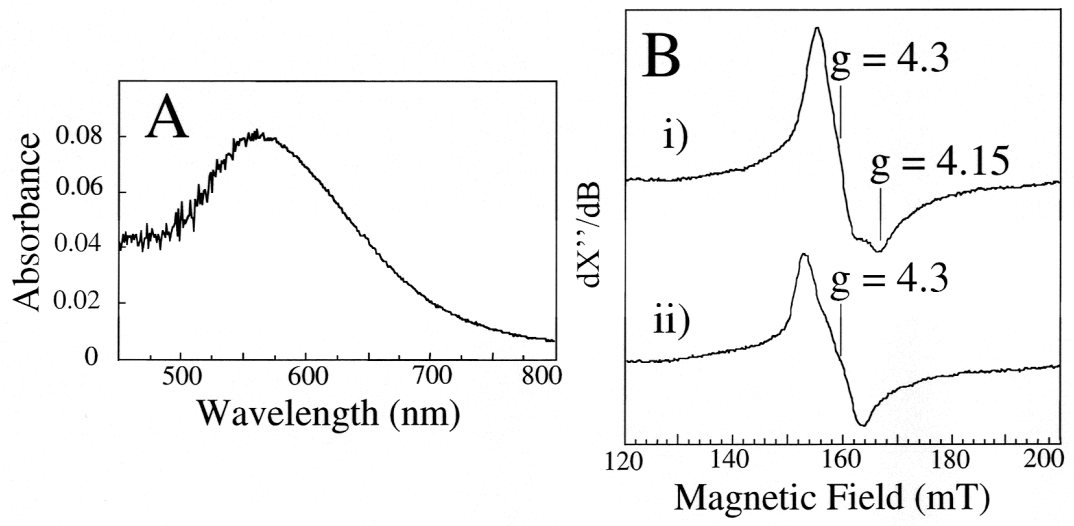

***Figure 1***. UV-visible (A) and X-band EPR spectra (B) of SOR E47A mutant from *D. baarsii* (200 μM in 50 mM Tris/HCl pH 7.6) treated with 6 equivalents $H_2O_2$ or 3 equivalents $K_2IrCl_6$. (A) UV-Visible spectrum recorded 5 s after addition of $H_2O_2$. (B) EPR spectrum after treatment with i) $H_2O_2$ and immediate freezing after mixing, ii) $K_2IrCl_6$. The contribution of the high-spin $Fe^{3+}$ center I $[Fe(SCys)_4]$ was subtracted from each UV-visible and EPR spectrum. EPR conditions : temperature 4.2 K, microwave frequency 9.676 GHz, power 20 mW, modulation 1.0 mT/100 kHz.

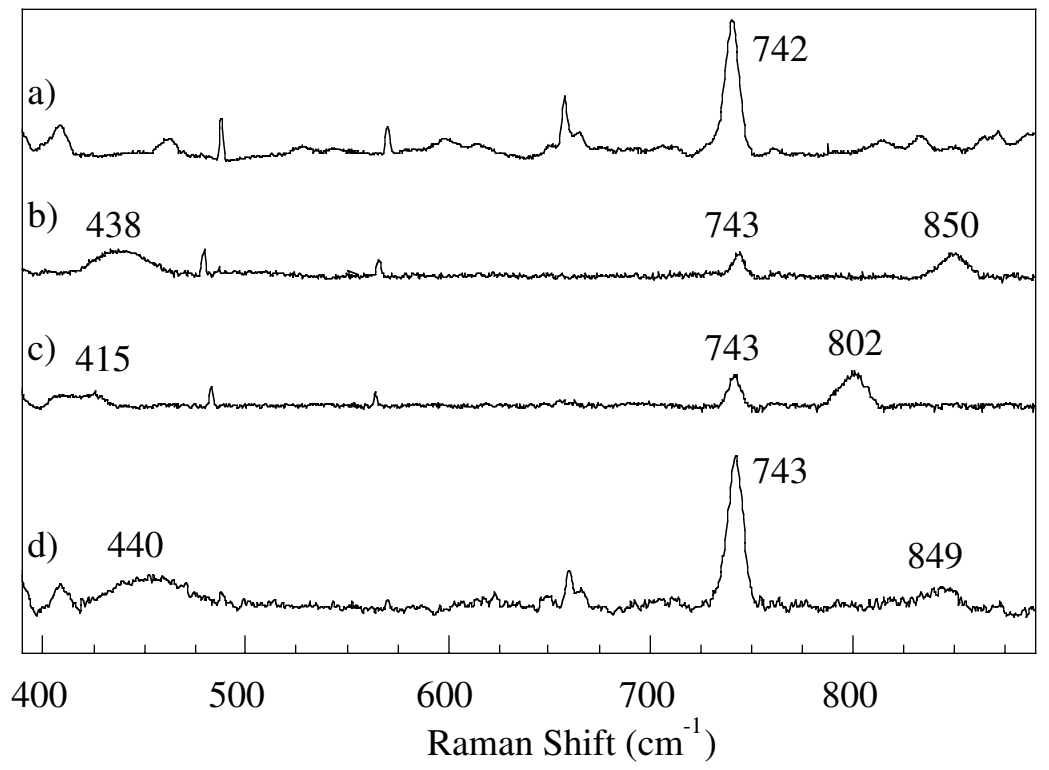

***Figure 2***. Resonance Raman spectra of SOR E47A mutant and wild-type forms from *D. baarsii* (1 mM in 50 mM Tris/HCl pH 7.6) excited at 647.1 nm (50 mW) at 15 K. a): SOR E47A treated with 3 equivalents $K_2IrCl_6$. b): SOR E47A treated with 6 equivalents of $H_2O_2$, rapidly mixed and immediately frozen (less than 5 s). c): SOR E47A treated with $H_2{}^{18}O_2$, same conditions as b). d): SOR wild-type treated with 6 equivalents of $H_2O_2$ rapidly mixed and immediately frozen (less than 5s).

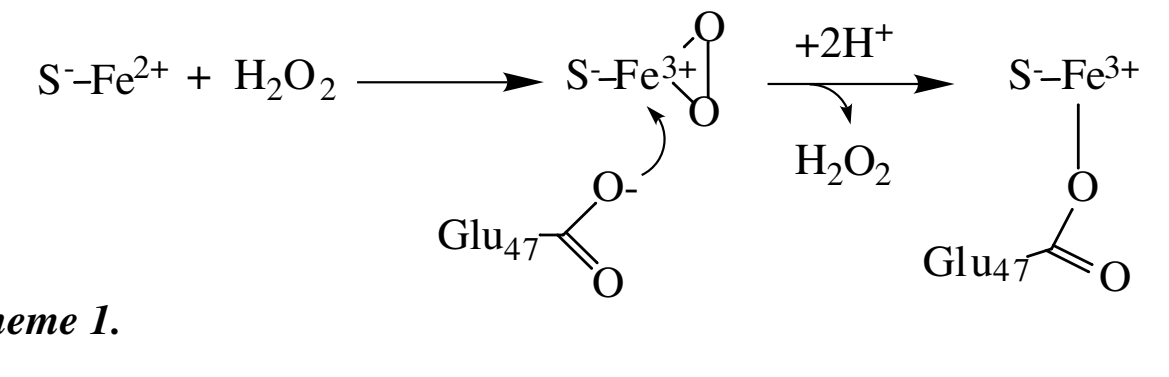

***Scheme 1.***

Table of Contents graphic :

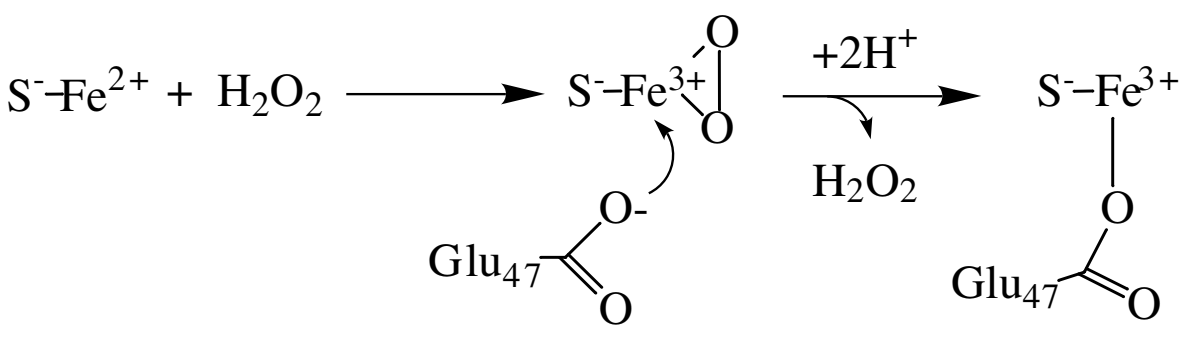

ABSTRACT FOR WEB PUBLICATION.

The active site of superoxide reductase SOR consists of a $Fe^{2+}$ center in an unusual [$His_4$ $Cys_1$] square pyramidal geometry. It specifically reduces superoxide to produce $H_2O_2$. Here, we have reacted the SOR from *Desulfoarculus baarsii* directly with $H_2O_2$. We have found that its active site can transiently stabilize a $Fe^{3+}$-peroxo species that we have spectroscopically characterized by resonance Raman. The mutation of the strictly conserved Glu47 into alanine results in a stabilization of this $Fe^{3+}$-peroxo species, when compared to the wild-type form. These data support the hypothesis that the reaction of SOR proceeds through such $Fe^{3+}$-peroxo intermediate. This also suggests that Glu47 might serve to $H_2O_2$ released during the reaction with superoxide.

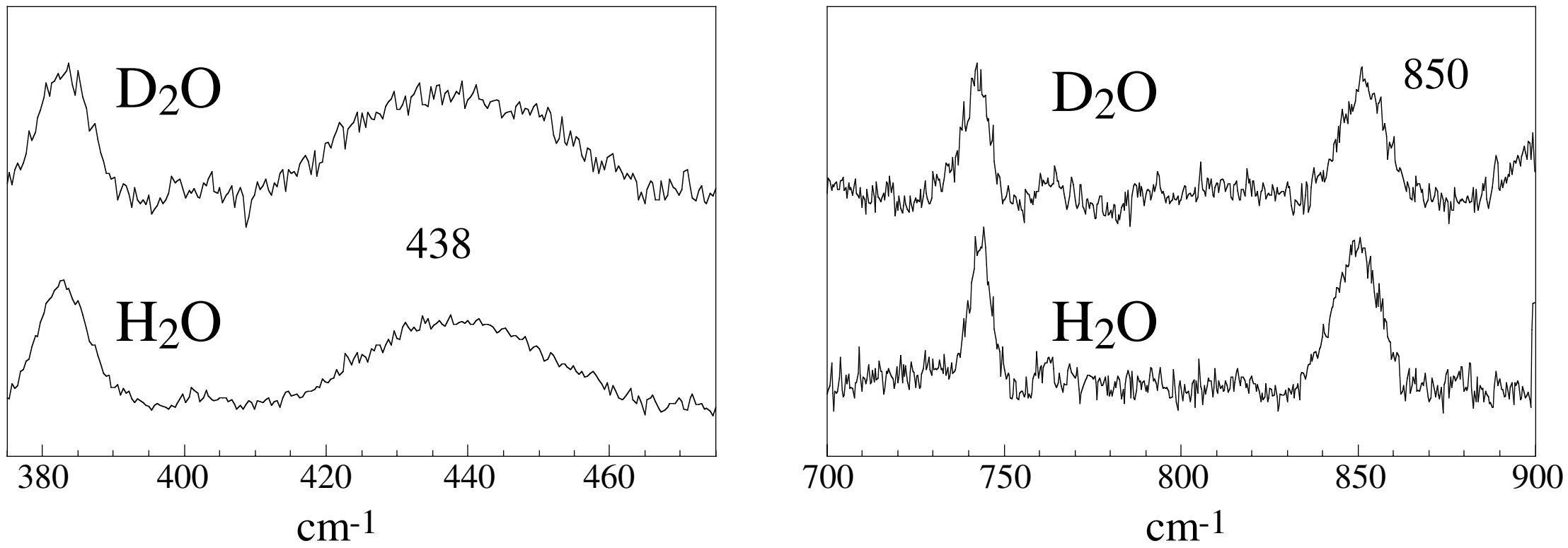

***Figure S1***. Deuterium isotopic effects on the resonance Raman spectra for the ν(Fe-$O_2$) (left panel) and ν(O-O) (right panel) regions of SOR E47A mutant from *D. baarsii* (1 mM in 50 mM Tris/HCl pH 7.6, or pD 8.0) excited at 647.1 nm (50 mW) at 15 K, treated with 6 equivalents of $H_2O_2$, rapidly mixed and immediately frozen (less than 5 s). Upper spectra in $D_2O$ solution. Lower spectra in $H_2O$ solution.